# Heterodimer nanostructures induced energy focusing on metal film

Ting Liu[a], Jingjing Hao[a], Yingzhou Huang*[,a], Xun Su[a], Li Hu[a] and Yurui Fang*[,b]

[a]Soft Matter and Interdisciplinary Research Center, College of Physics, Chongqing University, Chongqing, 400044, P. R. China

[b]Division of Bionanophotonics, Department of Applied Physics, Chalmers University of Technology, Gothenburg SE-412 96, Sweden

**Abstract**

As an interesting surface plasmon phenomenon discovered several years ago, electromagnetic field redistribution in nanoparticle dimer on film system provides a novel thought to enhance the light power on a plain film which could been widely used in surface enhanced Raman scattering (SERS), solar cells, photo-catalysis, etc. Homodimers on film are mainly investigated in past years, while the properties of heterodimers on film are still unclear. In this work, size difference induced electromagnetic field redistribution in Ag nanoparticle dimer on Au film system is investigated first. The results obtained from finite element method indicate that the smaller nanoparticle has much greater ability to focus light energy on Au film, which even reached more than 5 time compared to the larger one. Further researches indicate that this energy focusing ability has a strong relationship to the wavelength and diameter ration in dimer. Similar focusing phenomenon is found in the system of thick wire-smaller particle on film. Later, the SERS spectra collected in the small nanoparticle-large nanowire system provide an experimental evidence for this theoretic predication. Our results strengthen the understanding of surface plasmon on plane film and have potential application prospects in the surface plasmon related fields.

**Introduction**

Benefiting from the fast and great development of nanotechnology in the recent twenty years, plasmonics as an interdiscipline has achieved lots of interesting and profound progress in physics, chemistry and biology[1, 2]. The attractive merit of plasmonics is the ability of great electromagnetic field confinement at metal surface, which means bounding light within subwavelength area. The charming ability comes from the light excited collective oscillation of free electrons at metal surface, called surface plasmon polartions (SPPs). In a metal nanostructure, the confinement of light energy reaches maximal as the resonance of free electron collective resonance occurs, which is also called localized surface plasmon resonance (LSPR). Thanks to the huge enhanced light energy in some area at metal surface, there are numerous applications of LSPR reported in the past, such as surface enhanced Raman scattering (SERS)[3, 4], LSPR sensor[5, 6], surface catalysis[7, 8], hot electron generation[9, 10], harmonic generation[11, 12], solar cell[13, 14], etc.

Since the LSPR properties are greatly influenced by the size, shape, material and especially the plasmon coupling, various complex metal nanostructures has been investigated in the past, such as nanosphere[15], nanocube[16], nanorod[17], nanowire[18], nanorice[19], coreshell nanostructure[20], nanoarray[4], nanoparticle dimer[15], nanoparticle trimer[21], etc. Among them, nanoparticle dimers for the simple fabrication become the oldest and most common configuration of metal nanostructure to understand the physical mechanism of LSPR, especially in the study of SERS[15, 22]. However, most of previous SERS works has ignored the substrate influence on the LSPR of metal nanoparticle aggregates whereas all related SERS experiment was performed on substrate. In the last three years, we focused on this question and reported that an interesting electromagnetic field redistribution phenomenon was induced by the plasmon coupling between metal nanoparticle dimer and film (metal or dielectric), which have great significance in the SERS measurement of 2-dimension materials (grapheme, $MoS_2$, molecule monolayer, etc.) or other related fields[18, 23-25]. Therefore the research on plasmon coupling between nanoparticle aggregates and film is interesting and important topic in SERS and also in other related filed of plasmonics[26].

In this work, the LSPR properties of metal nanoparticle heterodimer on metal film are investigated. The nanoparticle heterodimer here consists of two nanoparticles with different sizes but the same material (Ag). This is because in the study of nanoparticle aggregates in plasmonics, there is always size difference in nanoparticles for the drawback of nanofabrication, especially for the nanoparticle synthesized by chemical methods, which is often neglected in nanoparticle dimer especially. Therefore, this work firstly investigates the distributions of electric field and surface charge in nanoparticle heterodimer on film through finite element method. The results indicate that there is an interesting energy focusing phenomenon occurs in the heterodimer-film system. In the system, the smaller nanoparticle confines much more light energy in the nanoparticle-film gap, which is greatly influenced by the size difference and wavelength of incident light. Furthermore, the results indicate that this size difference induced energy focusing phenomenon also occurs in nanoparticle-wire/film system, which is demonstrated experimentally here by the SERS spectra in Ag nanoparticle-nanowire hetrodimer on Au film.

## Experimental and theoretical section

### Simulation parameters

All simulation results in this work were obtained using the finite element method (COMSOL 4.3a commercial package). The heterodimer nanostructures here are consisted of two silver nanospheres (or silver nanosphere-nanowire) located 1nm above the gold film and the gap distance of heterodimer is set as 1 nm. The incident light illuminates from the heterodimer side normal to the substrate, whose polarization is along the heterodimer axis and the electric component is 1 V/m.

### Experimental section

The SERS sample of Ag nanoparticle-nanowire dimer on Au film were fabricated by the following method: the centrifuging washed Ag nanoparticles and Ag nanowires were immersed into $5\times10^{-6}$ M ethanol solutions of 2-Amino-5-nitrobenzenethiol (2A-5NBT) and Benzene-1,4-dithiol (BDT) molecules under magnetic stirring, respectively. 5 hours later, the Ag nanoparticle with 2A-5NBT solutions and Ag nanowire with BDT solutions were carefully centrifuging rinsed for 10 times and then mixed together. Through spin-coating method, the final solution was dropped on Au film and the Ag nanoparticle-nanowire heterodimer could be found through microscope.

The SERS measurements in this work were performed using a commercial Micro-Raman spectrometer (Horibba, LabRAM) with a 633 nm laser with polarization perpendicular to nanowire. The laser power illuminating the sample with a 100x objective in the experiments was measured at 2 mW.

## Results and discussion

The difference of LSPR properties between homogeneous dimer and heterodimer is firstly investigated and shown in Fig. 1. Here the homogeneous Ag dimers with both diameters of 80 nm and 160 nm on Au film are shown in Fig. 1a and 1b. The Ag heterodimer is consisting of two nanoparticles with diameters of 160 nm and 80 nm, respectively (Fig. 1c). The light with 633 nm wavelength propagates along normal direction of film surface illustrated by red arrow in Fig. 1(a), whose polarization is parallel to direction of nanoparticle dimer illustrated by yellow arrow. The electric field distributions in Fig. 1a-1c indicate that the electric field in nanoparticle-film gaps (L for the larger one while S for the smaller one in heterodimer, and L for left while R for right one in homogeneous dimer) were greatly enhanced in both homogeneous dimer and heterodimer situation, which are even much larger than that in nanoparticle-nanoparticle gap (P). This result is consistent with our previous work on electromagnetic field redistribution, which was caused by the plasmon coupling between surface charge on nanoparticle and induced image charge on film[18, 23-25]. Whereas, there are still obvious difference for electric field distribution in two systems that the electric field in gap of the smaller particle/film (gap S) is much stronger than that in gap of the larger particle/film (gap L) of heterodimer while they are the same in homogeneous dimer/film gaps . The inset table in Fig. 1a-1c show the exact value of electric field intensities in the gap excited at 633 nm wavelength light. Comparing the three configurations, one can see that the electric field in gap S is more than 2 times larger than that in gap L in heterodimer. However, the electric field in Gap L or R is the same in the homogenous dimer. Considering the light power is linear to $|E|^2$, this results meant more than 5 times light energy is confined by the smaller nanoparticle on film surface compared to the larger one. This finding is quite important in the plasmonic applications of nanoparticle dimer on film that the main contribution can come from the small nanoparticle since the size difference always exists in the real experiments. This phenomenon is very important in SERS measurement since the intensity of SERS is related to $|E|^4$. From the data in Figure 1c, it is very easy to deduce the obvious result that the Raman intensity from molecule at the bottom of small nanoparticle is more than 30 times larger than that of large one, which means the Raman signal enhanced by the large nanoparticle can be neglected. To understand the formation of this interesting energy focusing, the distributions of surface charges on nanoparticles and induced image charges on film surface are illustrated in Fig. 1a'-1c'. As in ref.25 illustrated, the induced image dipole always has a bonding configuration with the dimer dipole, so the opposite charges yield large enhancement in the gap for the homogeneous dimer/film system (Fig. 1a' and 1b'). For the heterodimer/film system (Fig. 1c'), the hybridization and induced image charge are similar. But there are two main aspects response for the energy focusing. First, the bigger particle has larger polarizability, which attracts more charges in the smaller one close to the gap. That means the interaction between the two particles will make the larger polarizability particle have a higher enhancement effect on the smaller one, and versus opposite. Second, the bigger particle has a larger distance with the induced imaging charge on the film which makes the field weaker, and larger radius of curvature will lower the field intensity because of the lightning rod effect. Besides, the charges on the bigger one are not as asymmetric as the smaller one, so the electric field in the closest to the film in the gap L is with both positive and negative signs while for the gap S, there are only the same sign of charges (Fig. 1f). The field distribution will vary at different resonance wavelength of hybridization, but the mechanism is similar, which will be discussed in the following.

To check if the smaller particle always has focusing effect in the whole spectral range, the wavelength dependent heterodimer/film system is also investigated as shown in Fig. 2. Here the simulated heterodimer/film system is the same as in Fig. 1c (Fig. 2a-2c). The images with three common Raman excitation wavelengths (532 nm, 633 nm and 785 nm) in Fig. 2a-2c intuitively indicate the variation of electric field distribution in the same nanoparticle-film system. To obtain more information, Fig. 2d illustrates absorption spectra of dimer systems with three configurations shown in Fig.1, in which the black, green and red line represented that homogenous dimer with 80nm, 160 nm diameters, heterodimer with 80nm and 160 diameters, respectively. There are more than two resonant peaks in spectra for all three configurations means the electromagnetic energy redistribution should vary at different resonance conditions. This data result in enhancement of electric field in all gaps varies a lot as the wavelength of incident light changes. However, although this electric field-wavelength relationships are quite different in three configurations, the field enhancement in gap S is always larger that gap L as shown in Fig. 2e. In all resonance conditions, the smaller particle exhibits energy focusing effect. And the enhancement in gap S is also better than that in gap P between the two particles in the main Raman measurement wavelengths after 532 nm (shown in Fig. S1). The much greater electric field in gap S (red solid line) presents two close peaks near 600 nm and the one in gap L (red dash line) presents two smaller peaks at 550 nm and 630 nm. Furthermore, the electric field in gap P decreases dramatically near 600 nm indicates the electromagnetic field redistribution is much more obvious in heterodimer compare to our previous work in homogenous dimer[25]. A comparison is also presented for the heterodimer and homogeneous dimers on film in Fig. 2e. We can see that the homogenous dimer with 80 nm diameter (black line) has stronger field enhancement at the longer wavelength resonant peak, but at the shorter resonant wavelengths, the enhancement for the heterodimer is always bigger than the homogeneous ones. It means that for the visible wavelength measurement, the heterodimer is always better than the homogeneous one that the energy focusing phenomenon is obviously exhibited in electric field intensities at the bottom of small nanoparticle (gap S) in three situations of nanoparticle dimer. To obtain further understanding of light energy distribution, the averaged electric field $E_{avg}$ is illustrated as a function of wavelength (red line for small nanoparticle and black line for large one) in Fig. 2f, where $|E_{avg}|^2 = \sum |E_i|^2 / S_{eff}$ represent the average light power focused by the nanoparticle on film. Here the $S_{eff}$ is area of circle with radius equal to 1/e decay length of electric field at the middle plane parallel to film in nanoparticle-film gap. Apparently, this analogical quality factor parameter shows that the small nanoparticle has a much greater ability to confine light on film, which reaches maximal near 600nm.

As illustrated in Fig. 2e, the size difference of nanoparticle in heterodimer is key factor to this energy focusing phenomenon. Therefore, the different diameter ratio with various wavelengths is also investigated in Fig. 3, in which other illuminating conditions remain as before. Here the diameter of small nanoparticle (represented by $D_S$) is fixed to 80 nm while the large nanoparticle (represented by $D_L$) varies (100 nm, 120 nm, 140 nm respectively). It is apparent that the induced energy focusing gets much stronger as the ratio of diameter increased in three images of electric field distribution with 633 nm wavelength. This can be understood by the surface charge distribution discussed above. In Figure 3a-3c, the electric field enhancement spectra in the three gaps for different dimer ratios are plotted. For the fixed small particle size, as the increasing size of the bigger particle, the field in the gap of bigger particle/film becomes weaker and weaker and the mode is also redshifted to middle infrared range. Meanwhile, the electric field enhancement in the smaller particle/film gap becomes stronger and stronger, and in a 7:4 size ratio, the field in the smaller particle gap is several times larger than that in the bigger particle/film gap. Since the light is also confined within the nanoparticle-nanoparticle gap (P), energy ratio R is illustrated as a function of wavelength to analysis the influence of size difference on light power enhancement on film surface in Fig. 3d, where $\mathbf{R}=|\mathbf{\mathit{E}}_s|^2/ (|\mathbf{\mathit{E}}_s|^2+|\mathbf{\mathit{E}}_l|^2+|\mathbf{\mathit{E}}_p|^2)$ represents the ratio of light confinement on film surface by small nanoparticle in all gaps of whole system. Apparently, the diameter ratio of heterodimer plays a great influence on the light confinement (R), where the black, green and red lines in Figure 3d represent the diameter ratio of 5:4, 6:4 and 7:4, respectively. Different R value and line type of three lines indicate not only the capability of light confinement by small nanoparticle but also the size difference determined confinement condition in wavelength. Furthermore, the apparent characteristic of this relationship is an obvious decrease in high energy region and increase in low energy region as the diameter ratio increasing. This can be explained that the plasmon coupling of two nanoparticle is strong in high energy region but weak in low energy region, which lead to the dominating light is confined within nanoparticle-nanoparticle gap (P) in high energy region but within small nanoparticle-film gap (S) in low energy region.

In practice, it is hard to decide the dimer direction even in dark field illumination, so it is very difficult to decide the exciting polarization. And it is also very hard to distinguish the nanoparticle-nanoparticle heterodimer with different molecule from the dimer with the same molecule in experiment because two kind of different molecules with close scattering cross sections need to be adsorbed on the two individual particle of the dimer respectively. Fortunately, the energy focusing effect also happens in hetero particle-wire / film system as well. A simulation in Fig. 4a and 4b show the theoretical results, which is very similar to the heterodimer/film system. In nanoparticle- nanowire system, the oscillation of SPPs at the nanoparticle-nanowire junction is similar to the nanoparticle-nanoparticle dimer when the polarization of incident light is perpendicular to the nanowire. Therefore, in all SERS measurement the polarization of 633 nm laser is always perpendicular to the nanowire in this work. The 3D distribution of electric field in nanoparticle adjacent to nanowire on film in Fig. 4a clearly illustrates the light is greatly confirmed at the bottom of nanoparticle or nanowire on film. Furthermore, the electric field intensities at adjacent part as a function of wavelength (solid line for and dash line for nanowire) in Fig. 4b indicate that the light energy ($|\mathbf{E}|^2$) is almost 4 times larger at the small particle bottom compare to that at large nanowire bottom at 633nm. This result also demonstrates that the smaller particle has strong focusing ability in a big range in the nanoparticle- nanowire system. However, the system with a bigger particle-thinner wire on film is different, as the energy will propagate on the wire so as to distribute on the whole wire, which is another issue we will work on.

To verify the theoretical predict discussed above, the SERS spectra of two different molecule with similar structure in Ag nanoparticle-nanowire on film are investigated whose Raman peaks are presented in Table 1 (the two kinds of molecules are with very close Raman scattering cross sections). As described in experimental section, the different molecule monolayer is absorbed on Ag nanoparticle or Ag nanowire, respectively (Fig. 4c, 2A-5NBT on particle and BDT on wire). The SERS spectra in Fig. 4c are collected at two typical positions in system that consisted of small Ag nanoparticle (about 94 nm diameter) adsorbed by 2A-5NBT molecule monolayer and lager Ag nanowire (about 120 nm diameter) adsorbed by BDT molecule monolayer is found in microscope. An obvious Raman peak at 1306 cm$^{-1}$ produced by 2A-5NBT molecule is presented in SERS spectrum collected at adjacent position of nanoparticle and nanowire (red line) but not in that collected at individual part of nanowire (black line) in Fig. 4c. To obtain further understanding of this difference, the SERS spectra are also collected from a similar individual Ag nanoparticle (about 122 nm diameter, red line) and no nanoparticle (black line) on the same Au film shown in Fig. 4d. For the Raman peaks at 1093cm$^{-1}$ or 1178cm$^{-1}$ generated by BDT molecule, the Raman intensities collected at adjacent position and individual nanowire are almost the same. However, for the Raman peak at 1336cm$^{-1}$ produced by 2A-5NBT molecule, the Raman intensity at adjacent position is more than 10 times larger than that at individual nanoparticle. This great difference in SERS enhancement of two molecule adsorbed at different Ag nanostructure demonstrates the Raman signals are mainly come from the molecule 2A-5NBT at small nanoparticle at adjacent position, which is contributed to the light energy focusing phenomenon of heterodimer that consisted of 94 nm diameter small nanoparticle and 120 nm large nanoparticle on metal film discussed above but not the electric field enhancement by nanoparticle on film. The results are reproducible in the smaller particle-larger wire on film system in a serious measurement. The results clearly show the light focusing effect we talked above.

## Conclusion

In this work, the diameter difference resulted electromagnetic field redistribution and energy focusing effect in nanoparticle heterodimer on metal film is investigated both theoretically and experimentally. The simulated results indicate that the small nanoparticle in dimer has a much greater ability to focus light energy on partilcle/film gap. Further studies of the electric field distribution in the heterodimer system indicate that the wavelength and the diameter ratio play important roles in this novel phenomenon. Furthermore, the SERS spectra collected in the small nanoparticle-large nanowire on film demonstrate this theoretic predication of the focusing ability of the smaller particles in the heterodimer/film system. Our results strengthen the understanding of the surface plasmon on plane film and have potential application prospects in the surface plasmon related fields e.g. SERS, solar cells, photocatalysis, etc.

## Author information

Corresponding Author
*E-mail: yzhuang@cqu.edu.cn. Phone: +86-15086879112.
*E-mail: yurui.fang@chalmers.se. Phone: +46-31-772 4540.

## Acknowledgement

This work was supported by the National Natural Science Foundation of China (11204390), Natural Science Foundation Project of CQ CSTC (2014jcyjA40002), Fundamental Research Funds for the Central Universities (106112015CDJXY300003) and Special Fund for Agro-scientific Research in the Public Interest (201303045).

Figures and Captions

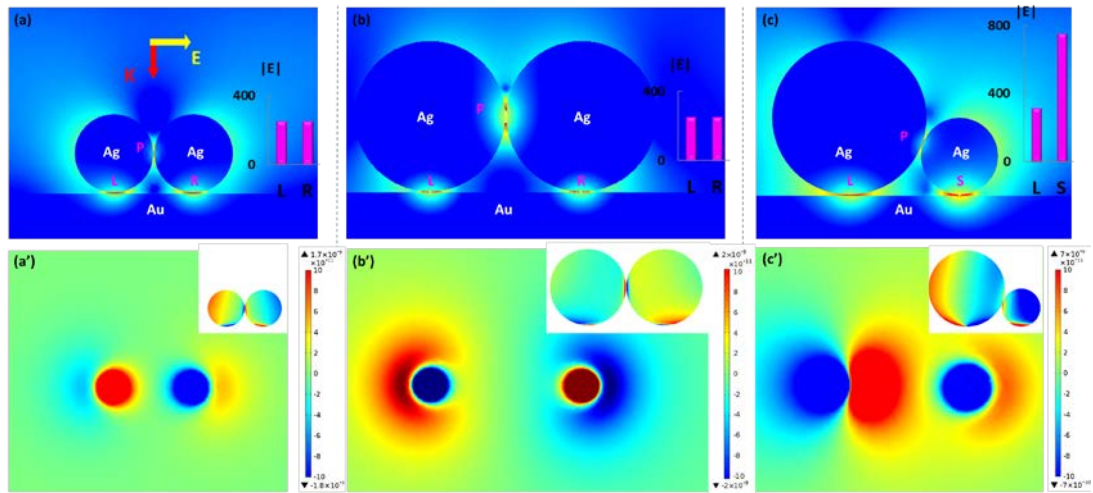

**Fig. 1** Electric field distribution of two Ag nanoparticle dimer on an Au film (a) two D = 80 nm nanoparticles, (b) two D = 160 nm nanoparticles, (c) D =160nm and D =80nm nanoparticles. The insets are the values of electric field intensities at gaps. (a'-c') Surface charge distributions on Au film and on Ag nanoparticles in Fig. (1a-1c). Here red arrow represents the propagation direction while the yellow one represents the polarization of incident light. The thickness of Au film is set as 100nm and all gap distances are set as 1nm.

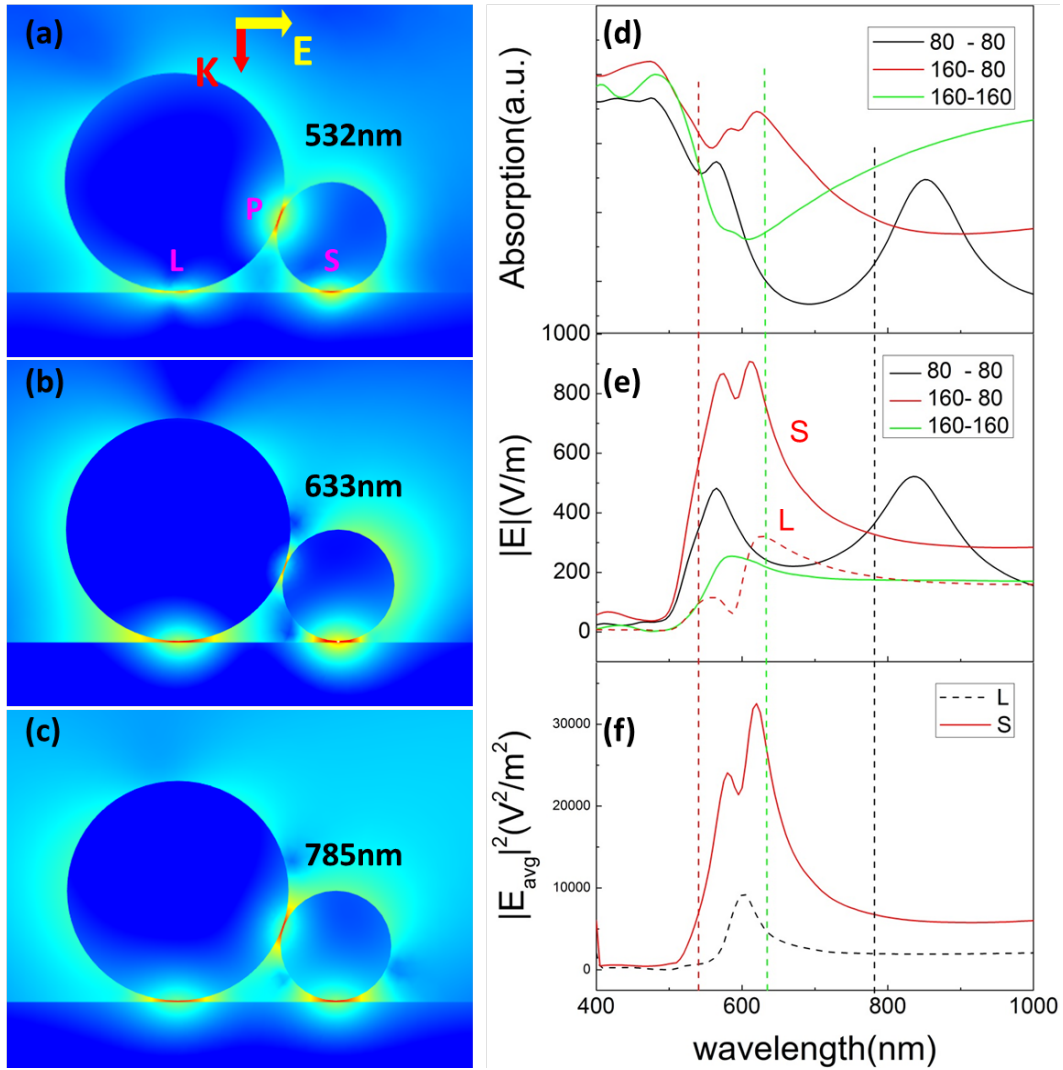

**Fig. 2** Electric field distribution of Ag nanoparticle heterodimer (D =160nm and D =80nm) on a 100 nm thick Au film with 1 nm gaps, the excitation wavelengths are (a)532 nm (b)633 nm, (c)785 nm. (d) Absorption spectra of three systems in Fig. 1. (e) Electric field intensities at gap S (solid line) and gap L (dotted line) of three systems in Fig. 1. (f) The $E_{avg}$ as a function of wavelength (solid line for small nanoparticle while dash line for large one). Here red arrow represents the propagation direction while the yellow one represents the polarization of incident light.

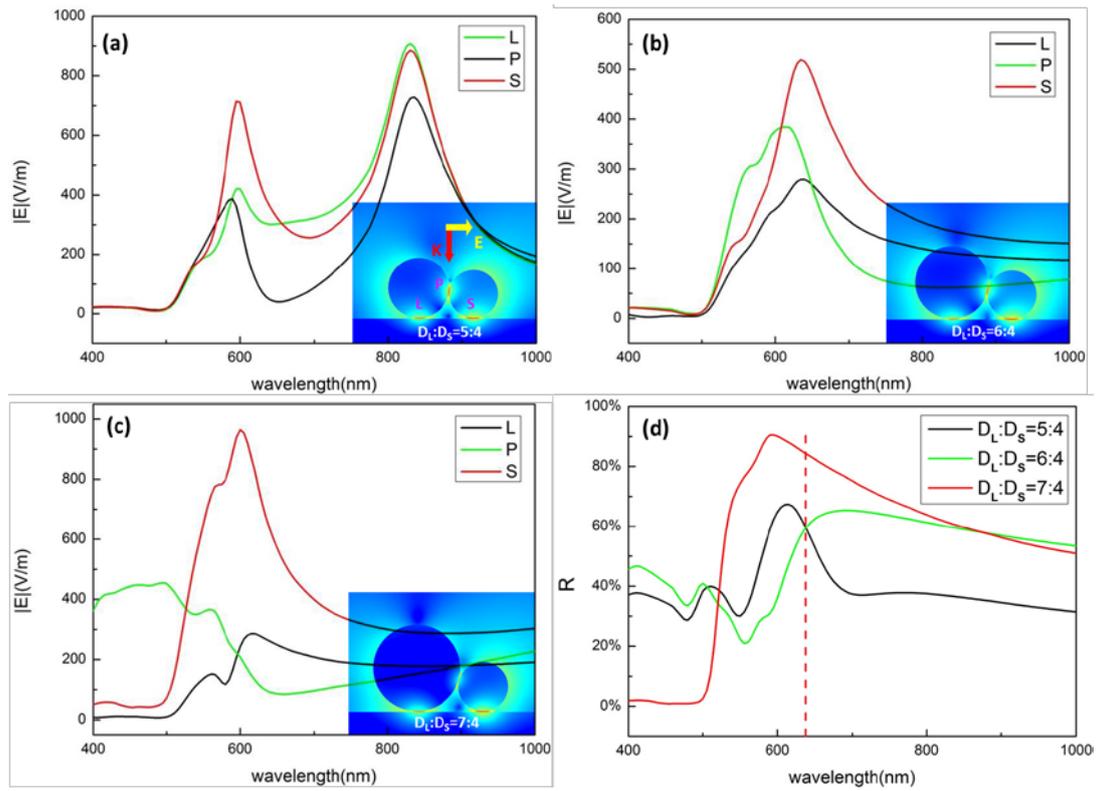

**Fig. 3** Electric field intensities at three gaps as function of wavelength (green line for gap L, black line for gap P and red line for gap S) in Ag nanoparticle heterodimer with different diameter ratio on Au film (a) $D_L:D_S = 5:4$, (b) $D_L:D_S = 6:4$, (c) $D_L:D_S = 7:4$. The insets are the corresponding electric field distributions. (d) The R as a function of wavelength in three systems in Fig. 3(a-c). Here red arrow represents the propagation direction while the yellow one represents the polarization of incident light.

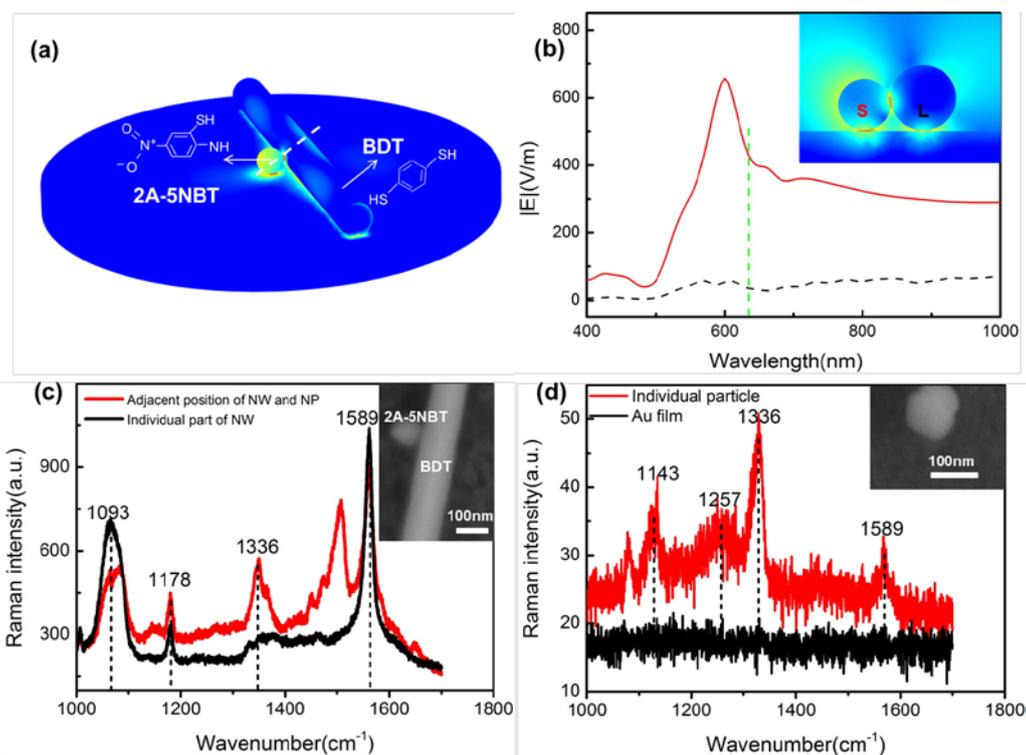

**Fig. 4** (a) 3D electric field distribution of D =94 nm Ag nanoparticle and D =120 nm Ag nanowire on a 100 nm thick Au film with 1 nm gaps. (b) Electric field intensities as a function of wavelength at gap S (solid line) and gap L (dash line) in Fig. 4a. the inset is the 2D electric field distribution. (c) SERS spectra collected at adjacent position of nanoparticle and nanowire (red line) and at individual part of nanowire (black line). (d) SERS spectra collected at an individual Ag nanoparticle (about 122 nm diameter, red line) and no nanoparticle (black line) on the same Au film.

| molecule | Typical Raman peaks (cm$^{-1}$) | | | | | |
|---|---|---|---|---|---|---|
| 2A-5NBT | X | 1143 | X | 1257 | 1336 | 1589 |
| BDT | 1093 | X | 1178 | X | X | 1589 |

**Table 1** The Raman peaks of 2A-5NBT, BDT molecules

# Supplemental Information

# Heterodimer nanostructures induced energy focusing on metal film

Ting Liu[a], Jingjing Hao[a], Yingzhou Huang*[,a], Xun Su[a], Li Hu[a] and Yurui Fang*[,b]

[a]Soft Matter and Interdisciplinary Research Center, College of Physics, Chongqing University, Chongqing, 400044, P. R. China

[b]Division of Bionanophotonics, Department of Applied Physics, Chalmers University of Technology, Gothenburg SE-412 96, Sweden

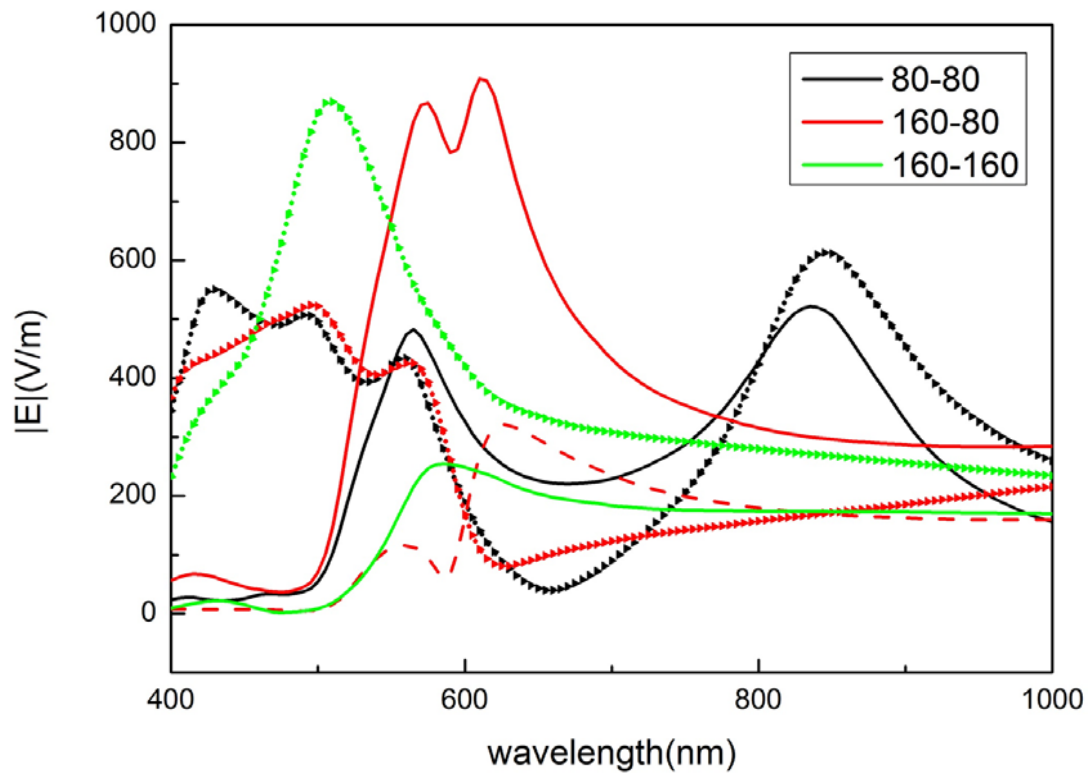

**Figure S1** | Electric field intensities at nanoparticle-nanoparticle gap (P, triangle line), small nanoparticle-film gap (S, solid line) and large nanoparticle-film gap (P, dash line) as a function of wavelength in three typical systems including two D = 80 nm nanoparticles (black line), two D = 160 nm nanoparticles (green line), a D =160nm and a D =80nm nanoparticles (red line).